\documentclass[twocolumn]{aastex631}
\usepackage{newtxtext,newtxmath}
\usepackage{amsmath,amssymb}	

\begin{document}

\title{Investigation of the Gamma-Ray Bursts prompt emission under the relativistically expanding fireball scenario}

\correspondingauthor{Soumya Gupta, Sunder Sahayanathan}
\email{soumya.gupta1512@gmail.com/soumyag@barc.gov.in, \\sunder@barc.gov.in}

\author[0000-0001-6621-259X]{Soumya Gupta$^{*}$}
\affiliation{Homi Bhabha National Institute, Mumbai, Maharashtra, India}
\affiliation{Bhabha Atomic Research Center, Mumbai, Maharashtra, India}

\author{Sunder Sahayanathan$^{+}$}
\affil{Homi Bhabha National Institute, Mumbai, Maharashtra, India}
\affil{Bhabha Atomic Research Center, Mumbai, Maharashtra, India}

\begin{abstract}
The spectral properties of a composite thermal emission arising from a relativistic expanding 
fireball can be remarkably different from the Planck function. We perform a detailed study
of such a system to explore the features of the prompt emission spectra from the gamma-ray 
bursts (GRBs). Particularly, we address the effect of optical opacity and its dependence on 
the density profile between the expanding gas and the observer. This results in a nontrivial 
shape of the photospheric radius which in combination with the constraints derived from the
equal-arrival-time can result in a mild broader spectrum compared to the Planck function.
Further, we show the time-integrated spectrum from the expanding fireball deviates significantly
from the instantaneous emission and is capable of explaining the observed broad spectral width of 
the GRBs. We also show, that the demand of the spectral width of the order of unity, obtained through
statistical analysis, is consistent with the scenario where the dynamics of the expanding fireball 
are governed predominantly by the energy content of the matter.

\end{abstract}


\section{Introduction}\label{sec:intro}
Gamma-Ray Bursts (GRB) are the brightest events that spate the radiation in a brief duration of time. 
They carry tremendous energy lasting from a few ms to a hundred seconds, which outshines any sources 
in the sky. This short-lived burst of radiation, bright in $\gamma$-ray and hard X-ray, is referred 
to as the prompt phase of GRB. Even after decades of study using several space-based missions, 
the radiation mechanism of the prompt phase is still under debate \citep{rees_2005,peer_08,beloborodov_2011,zhang14}. 
Among various models, the competitive ones are the decoupled thermal emission associated with the 
expansion of an initial optically thick fireball \citep{cavallo_78,paczy_1986,Goodman_86,shemi_90,piran_shemi_naray} and the non-thermal origin due to synchrotron loss 
from an electron distribution accelerated at the shock fronts \citep{rees94,Tavani1996,cohen_1997,sari_1997,Bonjak2009,daigne_2011,zhang11}. Nevertheless, under the 
fireball scenario, both these emission products are viable since after the release of the initial 
thermal energy, the matter coasts relativistically developing shocks which dissipate the bulk energy 
of the flow \citep{Thompson_1994,ghisellini99,lazzati_2010,lazzati_2011,Beloborodov12}.

The dynamics of an expanding fireball are mainly governed by the factor $\eta$, the ratio of the energy 
content in radiation $E_0$ to the total rest mass energy of the matter $M_0 c^2$ at the beginning of the 
burst \citep{shemi_90,piran_shemi_naray}. In the initial phase of evolution, the fireball is optically thick and expands freely at 
the expense of its internal energy ($\eta\gg1$),
with the bulk Lorentz factor ($\Gamma$) of the flow proportional to the radius ($\Gamma \sim R$) \citep{Goodman_86}. 
The temperature of the fireball at this phase will largely remain constant. The acceleration phase 
continues till $\eta \sim 1$ and from there on, the energy content of the matter dominates the evolution 
of the fireball. During this matter-dominant phase ($\eta\ll 1$), the fireball coasts with a constant 
bulk velocity and the temperature scales as $R^{-2/3}$ \citep{piran_shemi_naray}.  

In the course of evolution, the temperature of the fireball drops and once it falls below the pair 
production threshold, the radiation decouples from the matter, releasing a burst of radiation. The 
radius of the fireball where the last scattering surface of the trapped photons falls is referred 
to as photospheric radius $R_{ph}$.Interestingly, $R_{ph}$ can fall either during $\eta>1$ or $\eta<1$ 
phase and the observed burst property will vary significantly under these two scenarios. When $\eta>1$, 
the matter is still in the accelerating phase and the instantaneous Doppler boosting nullifies 
the adiabatic loss. Hence, the observed burst luminosity will be equal to that of the initial outflow \citep{meszaros_2000}. 
On the other hand, when $\eta < 1$, the flow incurs significant adiabatic loss and the observed burst 
luminosity will be less than the initial outflow.

The effect of the viewing angle and the relativistic flow have a substantial role in the observed 
shape of the photospheric surface of the burst. For a spherically symmetric expanding matter, the 
photospheric radius measured by a distant observer $ R_{ph}(\theta)$ will be larger for the higher 
viewing angle (high latitude) of the emission cone. This will result in a non-trivial concave shape 
of the photospheric surface (\S\ref{ep})(\cite{peer_08}, \cite{meng2018}). Similarly, due to the 
relativistic expansion, the observed instantaneous emission released from a certain radius has to 
travel a long distance in case of higher latitude compared to the on-axis emission. Hence, the surface 
constructed from the photons reaching the observer at the same time ({\it equal-time-surface }) will deviate 
from a spherical shape. If the expansion is associated with a fall in the temperature of the fireball, 
then the {\it equal-time-surface } will be a superposition of multiple temperatures. These effects will 
significantly modify the initial observed emission profile and the spectrum of the GRB prompt 
phase (\S \ref{os}).   

Irrespective of the radiative model, the prompt emission of GRBs is often well represented by an 
empirical function involving a broken power-law with an exponential turnover, commonly referred 
to as the Band function \citep{band93}. This function is useful to identify the peak frequency, the low 
energy and the high energy spectral indices which can be readily compared with the constraints 
derived from the radiative models \citep{preece_2002,peer_zhang_2006,uhm_14}. Besides these, it also helps to understand the statistical 
property of a large sample of GRBs, for example, the correlation between the peak energy and the 
isotropic luminosity \citep{amati02,yonetoku04,liang10}. Another study involving a large sample of GRBs is their 
spectral width (FWHM: $\mathcal {W}$) distribution derived from the best-fit Band function parameters. 
For long GRBs the most probable $\mathcal {W}$is found to be $\sim 1.07$ estimated from a sample 
of 1873 GRBs and in the case of short GRBs, it was $\sim 0.86$ obtained from a sample of 419 
GRBs \citep{axelsson_width}. Interestingly, these widths are much broader than that of a Planck function 
(FWHM $\approx$ 0.54) while much narrower than a typical synchrotron spectrum (FWHM $\sim$ 1.6).

In an attempt to explain the unusual $\mathcal {W}$ of GRBs, \cite{sahayanathan} considered 
a multi-temperature blackbody spectrum associated with an evolving fireball. Nevertheless, 
the $\mathcal {W}$ of $\sim$ 1 demanded an unphysical faster rate of temperature evolution 
and this disfavoured the fireball interpretation of GRB prompt emission. In this work, we 
revisit the expanding fireball scenario of GRB prompt emission and show that a proper inclusion
of the radial evolution is capable of explaining the observed $\mathcal {W}$ of GRBs. The 
effect of the concave photospheric surface and the emission cone limited by the relativistic 
effects are also considered in studying the temporal evolution of the fireball blackbody spectrum.

\section{EXPANDING FIREBALL}\label{sec_fireball}
\subsection{Evolutionary Phase}\label{ep}
\begin{figure}
    \centering
    \includegraphics[width = 0.32\textwidth]{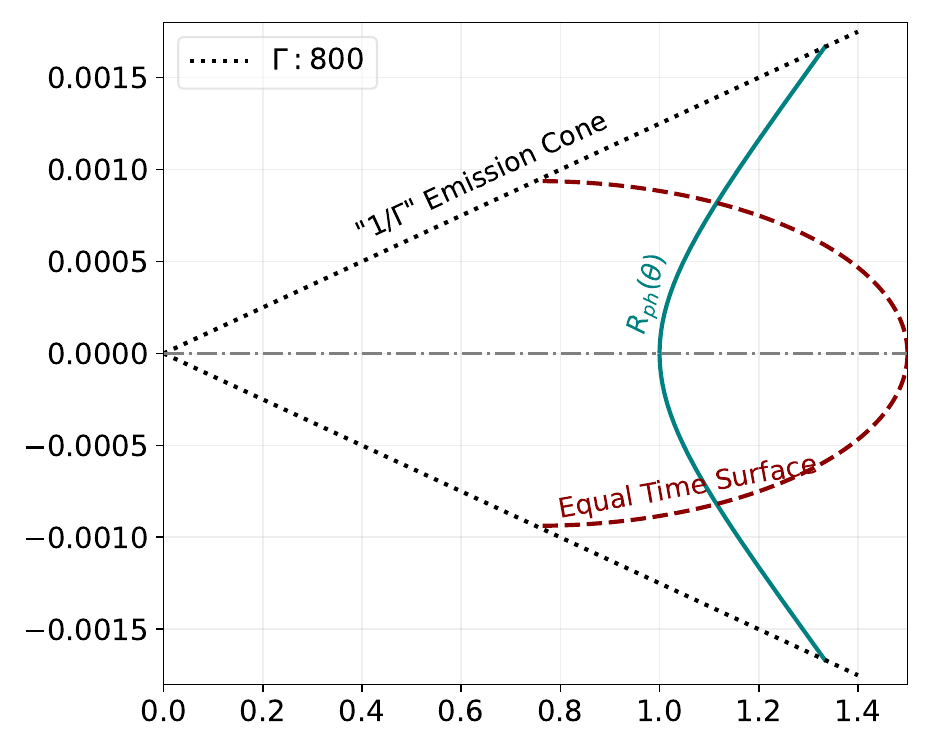}
   \caption{The plot depicts the {\it equal-time-surface} in red-dashed line, photospheric 
   radius $R_{ph} (\theta)$ in teal-solid line and emission cone of angle $1/\Gamma$ in 
   black-dotted line for the $\Gamma$ of 800. The $R_{ph0}$ is scaled to unity and the 
   grey-dot-dashed line is the axis for $\theta=0$.}
    \label{fig_rph}
\end{figure}

The optical depth of a relativistically expanding spherical symmetric wind 
measured for a photon propagating 
towards an observer will be \citep{abramowic1991}
\begin{align}
	\tau(r,\theta) = \frac{r_d}{\pi r}(\theta -\beta \sin\theta)
\label{tau_e1}
\end{align}
where, $r$ is projected distance and $\theta$ is the angle with respect to the line of 
sight such that 
$r\,\sin \theta$ is the radial distance $R$ of the emitted photon from the 
centre of the spherical wind. The quantity $\beta$ is the velocity of expansion of 
the wind, assumed to be constant and expressed in the units of $c$,
and $r_d$ is the radius of the opaque disk measured by the observer for photons
emerging from infinity. The observed photospheric radius will then be \citep{peer_08}
\begin{align}
    R_{ph}(\theta) = \frac{R_{ph0}}{(1-\beta)}\left(\frac{\theta}{\sin \theta}-\beta\right)
\label{R1}
\end{align}
where, 
$R_{ph0} \approx r_d/2\pi \Gamma^2$
is the minimum photospheric radius measured at $\theta=0$ and
$\Gamma$ is the bulk Lorentz factor of the plasma. Interestingly,
$R_{ph}$ will be an increasing function of $\theta$ and hence, the 
observer will visualize the photospheric surface as concave though
the expansion is spherically symmetric (Figure \ref{fig_rph}).

Due to spherical expansion, the photons emitted on-axis will arrive 
earlier than those emitted at higher latitudes. Hence, the photons 
received by an observer at an instant will be a combination of 
higher latitude emissions, when the expanding plasma is young relative
to the on-axis emission. The evolutionary {\it equal-time-surface} corresponding
to the instantaneous emission by the observer can then be obtained as \citep{sahayanathan}
\begin{align}
	R(\theta)&= R(0)\left(\frac{1-\beta}{1-\beta \cos\,\theta}\right)
    \label{ets1}
\end{align}
which is a decreasing function of $\theta$ and will be an ellipsoidal surface (Figure \ref{fig_rph}).
Due to relativistic expansion, the emitted photons will be confined within a cone of 
a semi-vertical angle $1/\Gamma$. Hence, this angle limits the maximum higher latitude 
emission received by the observer with $R(\theta)|_{\rm max}\approx R(0)/(1+\beta)$.

During the initial stage of evolution, the opening angle can be significantly 
be smaller than $1/\Gamma$ and will be governed by $R_{ph}(\theta)$.
Since $R(\theta) \geq R_{ph}(\theta)$ and from 
equations \ref{R1} and \ref{ets1} we obtain
\begin{align} 
	\frac{r_d}{\pi R(0)(1-\beta)}\left(\frac{\theta}{\sin\theta}-\beta\right)(1-\beta\cos\theta)\leq 1
    \label{trance}
\end{align}
For a given $R(0)$, the opening angle of the emission cone can be obtained
by solving the above transcendental equation which will be smaller than $1/\Gamma$. 
The on-axis radial distance beyond which the angle of the 
emission cone will be constant at $1/\Gamma$ will be $R(0)\gtrsim (1+\beta) R_{ph0}$.

{As the fireball expands beyond $R_{ph}(\theta)$, the photons decouple 
from the inner regions and have a finite probability to be scattered along the 
line of sight \citep{peer_08,peer_2011,beloborodov_2011}.
The observed intensity of the radiation will then be proportional to  
\begin{align}
    \zeta(\theta)&=\int\limits^{R(\theta)}_{R_{in}}\exp[{-\tau(z,\theta)}] dz
	\label{zeta_1}
\end{align}
where, the direction of $z$ is along the line of sight and 
$R_{in} = {\rm MIN}\{[R(\theta)-R_{ph}(\theta)],R_{ph0}\,\cos(\theta)\}$. Using equations \ref{tau_e1} and \ref{R1}, we obtain
\begin{align}
	\zeta(\theta)= \int\limits^{R(\theta)}_{{R_{in}}} \exp\left[{\frac{-\cos(\theta)\,R_{ph}(\theta)}{z}}\right] dz 
	\label{zeta_2}
\end{align}
}

The observed temperature of the fireball evolves depending on the energetics of
the burst phase. In the radiation dominated phase ($\eta \gg 1$), $T_{\rm obs} \sim {\rm constant}$;
while, in the matter dominated phase ($\eta \ll 1$), $T_{\rm obs} \propto R^{-2/3}$.
If we represent the on-axis temperature of the fireball as $T_0$, then the 
temperature gradient observed over the {\it equal-time-surface } will be 
\begin{align}
	\frac{T_{\theta}}{T_0}&=\left[\frac{R(0)}{R(\theta)}\right]^{\alpha} \nonumber \\
&=\left(\frac{1-\beta \cos\,\theta}{1-\beta}\right)^\alpha
    \label{t01}
\end{align}
where, $\alpha$ varies from 0 to 2/3. In terms of the on-axis photospheric temperature 
$T_{ph0}$, the temperature gradient can be expressed as 
\begin{align}
	\frac{T_{\theta}}{T_{ph0}}=\left(\frac{R_{ph0}}{R(0)} \frac{1-\beta \cos\theta}{1-\beta}\right)^\alpha
    \label{t1}
\end{align}

\subsection{Observed spectrum}\label{os}

The instantaneous flux from the expanding fireball measured by an observer 
located at a distance $D_L$ will be
{
\begin{align}
	F_\nu&=\frac{2\pi}{D_L^2}\,\int\limits^1_{\cos \phi} I_\theta(\nu)\,R(\theta)^2\,\zeta(\theta)\, \mu d\mu  \nonumber \\
    & = \frac{2\pi R(0)^2 (1-\beta)^2}{D_L^2} \int\limits^1_{\cos \phi} \frac{I_\theta(\nu)}{(1-\beta\mu)^2}\,\zeta(\theta)\, \mu d\mu 
	\label{fnu}
\end{align}
} 
where, $I_\theta(\nu)$ is the observed specific intensity and $\mu = \cos\theta$. Using Lorentz invariance, $I_\theta(\nu)$
can be expressed in terms of the expanding gas comoving specific intensity $I^{'}_\theta(\nu^{'})$ 
as
\footnote{All the quantities represented with $'$ are measured in the expanding gas comoving frame.}
\begin{align}
	I_\theta(\nu) &=	I^{'}_\theta(\nu^{'}) \left(\frac{\nu}{\nu^{'}}\right)^3 \nonumber  \\
	&=\frac{2h}{c}\frac{\nu^{3}}{\exp{\left(\frac{h\nu}{kT_{\theta}}\right)}-1} 
    \label{fnu_1}
\end{align}	


\begin{figure}
	\centering
    \includegraphics[width = 0.32\textwidth]{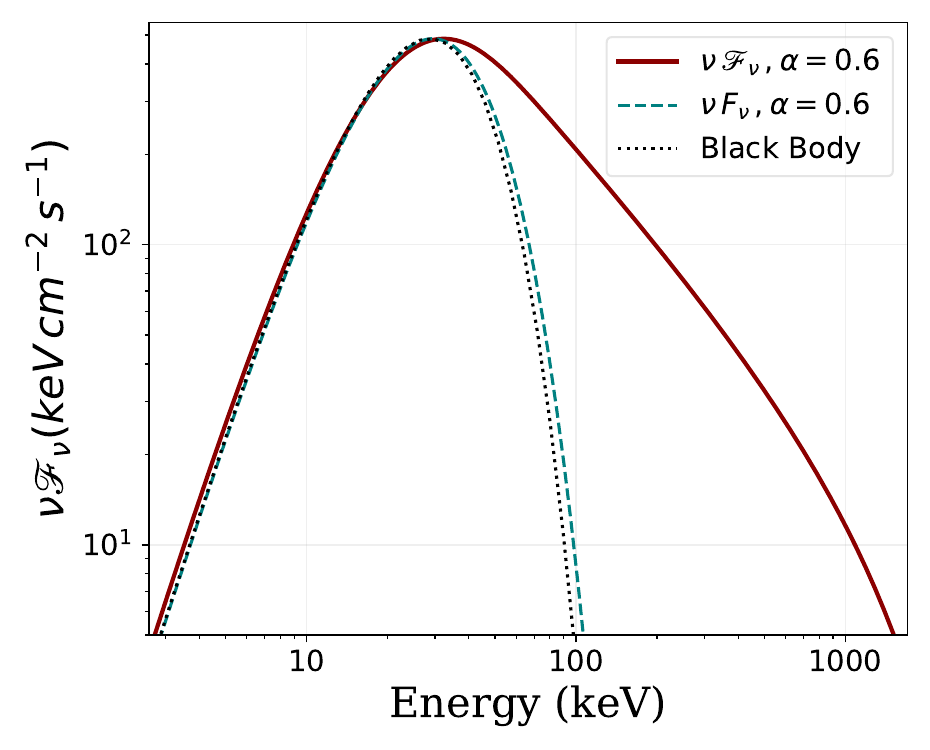}
    \caption{The plot represents the spectrum ($\alpha=0.6$) for the time-averaged case ($\nu\,\mathscr{F}_\nu$) in a 
    red-solid line, the instantaneous case ($\nu\,F_\nu$) in an teal-dashed line and the 
    black body in a black-dotted line.
    }
    \label{fig_spec}
\end{figure}
\begin{figure}
	\centering
    \includegraphics[width = 0.35\textwidth]{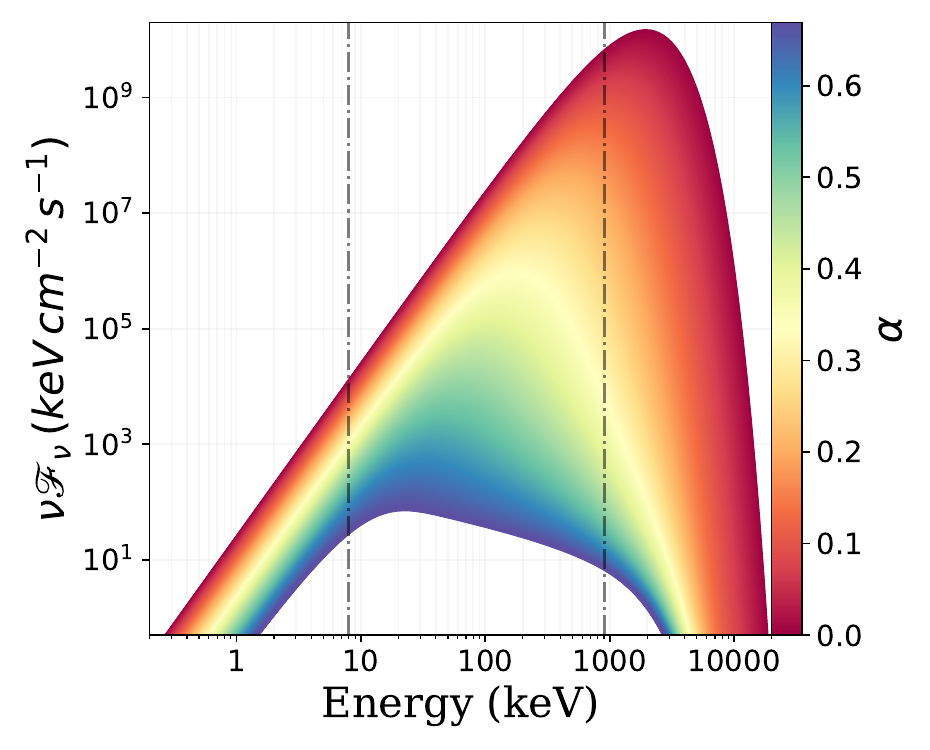}

    \caption{
    The plot represents the change in the time-integrated spectrum for the $\alpha$ 
    ranging from 0 to $2/3$, color-mapped on the right axis.
    {The vertical grey-dashed line represents the energy range of 
    the NaI detector of the Gamma-Ray Burst Monitor (GBM) onboard \textit{Fermi}.}}
    \label{fig_alp}
\end{figure}


and using equations \ref{zeta_2}, \ref{fnu} and \ref{fnu_1}, the observed instantaneous flux will be 
{
\begin{align}
	F_\nu&= 4\pi h \nu^3\left[\frac{(1-\beta)R(0)}{cD_L}\right]^2 \times \nonumber \\
	& \int\limits^1_{\cos\phi}\frac{\mu }{(1-\beta\mu)^2} 
	\frac{\zeta(\theta)\,d\mu}{\left\{\exp\left[\frac{h\nu }{kT_{ph0}}\left( \frac{R(0)}{R_{ph0}}\frac{1-\beta}{1-\beta\mu}\right)^\alpha \right]-1\right\}}
    \label{fnu_2}
\end{align}}
In {Figure \ref{fig_spec}}, we show the instantaneous spectrum from the fireball along with an 
equivalent Planck function for $\alpha$ ranging from 0 to 2/3. The temperature variation will be more pronounced for the time-averaged spectrum and will
significantly deviate from the Planck function. Let us assume the evolution of the fireball
for a duration $\Delta t$ during which the on-axis radius expands from $R_1$ to $R_2$ ($= R_1 + 2 \beta c\Gamma^2 \Delta t $). 
The time-averaged spectrum during  
$\Delta t$ will be equivalent to 
\begin{align}
    \mathscr{F}_\nu = \frac{\int\limits^{R_2}_{R_1} F_\nu(x)\, dx}{R_2-R_1} 
        \label{fnu_t}
\end{align}
In {Figure \ref{fig_spec}}, we show the time-averaged spectrum for a duration of 5s along with the 
Planck function. The spectral shape has a strong dependence on $\alpha$ (in Figure \ref{fig_alp}) and in Figure \ref{fig_var}, 
we show the variation in the peak of the spectrum with respect to 
various model parameters. To study the variation in the spectrum due to an individual parameter,  
the rest of the parameters were fixed at $\Gamma = 100$, $R_{ph0} = 12.0$ log cm, 
$T_{ph0} = 500$ keV, duration of burst {$\Delta t$} = 5.0 sec and $\alpha = 0.67$ (Figure \ref{fig_alp}, \ref{fig_var} and \ref{fig_width}). We note the flux at the spectral peak enhances 
nearly by an order of ten (in Figure \ref{fig_alp}), when $\alpha$ is varied from 0 to 2/3. However, the dynamic range of the 
detector will typically be of the order of 2-3. Therefore, the range of $\alpha$ that could be studied by an instrument will be $\Delta\alpha \sim 0.15$.


\subsection{Spectral width}\label{sw}
\begin{figure}
    \centering
    \includegraphics[width=0.45\textwidth]{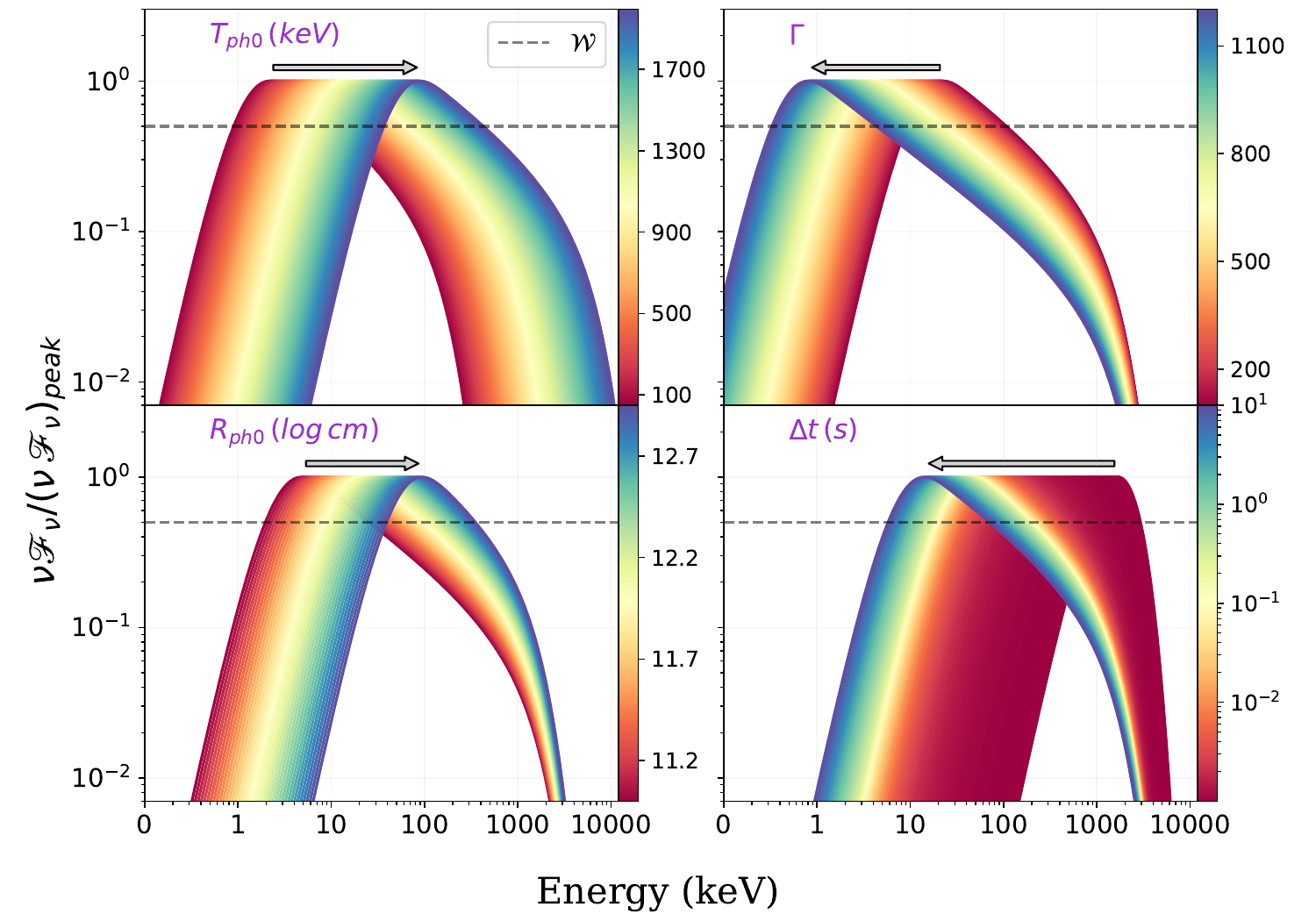}
    \caption{The plot exhibits the evolution of the time-integrated spectrum with different 
    parameter space. The variation with $T_{ph0}$, $\Gamma$, 
	$R_{ph0}$ and $\Delta t$ are plotted in the {top-left, top-right, bottom-left 
	and bottom-right panels} respectively, and color-mapped on the right axis of each panel. 
	The grey-dashed line is the spectral width.  The spectrum is scaled with the value at peak 
	energy to illustrate the shift in peak energy (direction of the arrow) with an increase 
	in model parameters. }
    \label{fig_var}
\end{figure}

The time-averaged spectrum from the fireball can imitate various spectral shapes with the proper
choice of $\alpha$. Accordingly, their spectral width at half the maximum flux ($\mathcal {W}$, defined as
$\mathcal {W}$=log($\nu_2$/$\nu_1$), can vary with $\alpha$. Besides this, the variation in other parameters may also 
impact $\mathcal {W}$. To investigate this, we estimated $\mathcal {W}$ for a wide range of model parameters
and the results are shown in Figure \ref{fig_var}. It is evident the variation in $\mathcal {W}$ is not prominent in the case of 
the parameters $\Gamma$, $R_{ph0}$, $\Delta t$ and $T_{ph0}$ (The flux at half maximum is shown as a grey-dashed line in Figure \ref{fig_var}). However,
significant change in W can be witnessed even with moderate variation in $\alpha$. 

\begin{figure}
	\centering
    \includegraphics[width = 0.45\textwidth]{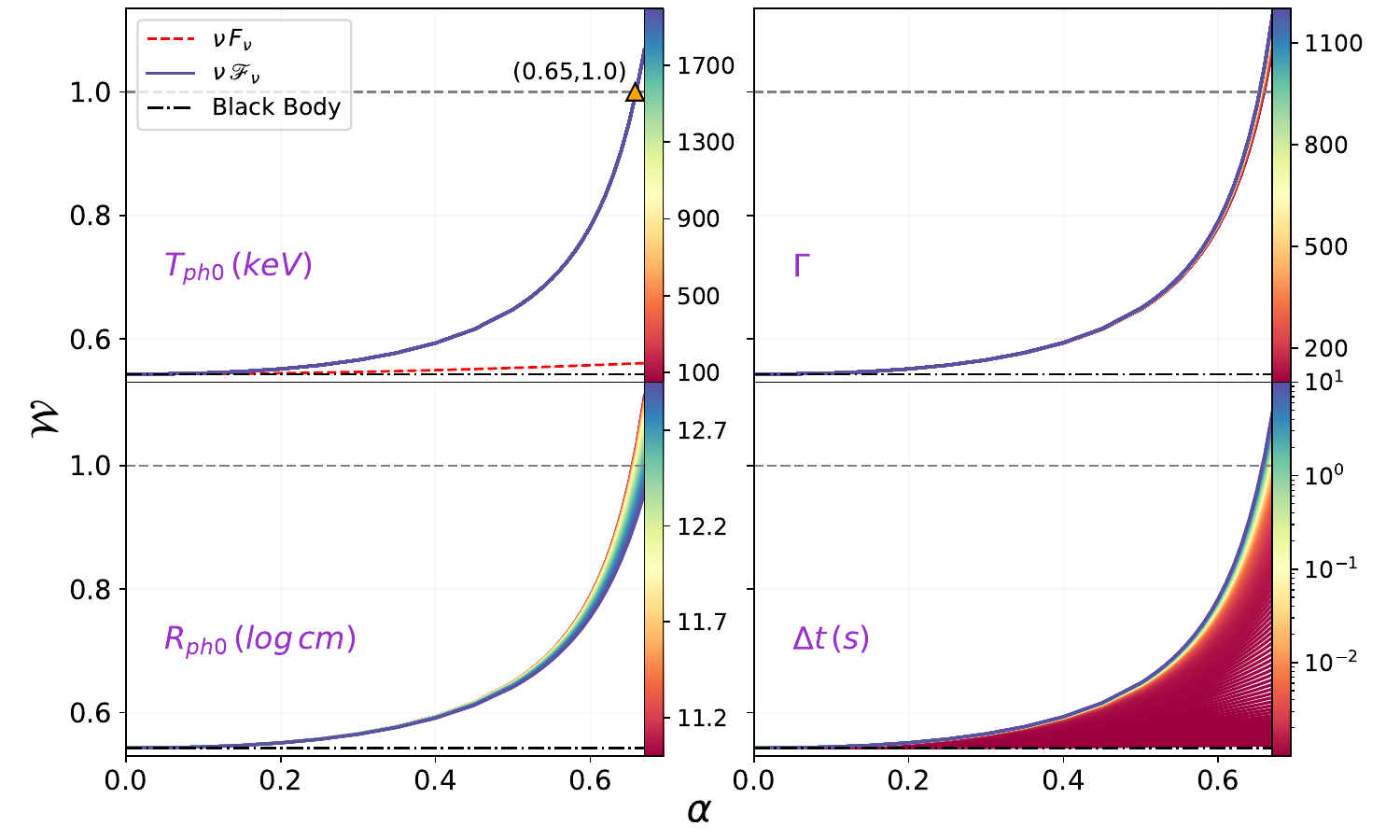}
    \caption{{The plot represents the variation of $\mathcal{W}$ of time-averaged spectra ($\nu\,\mathscr{F}_{\nu}$) with $\alpha$, for different 
    parameter space. The variation with $T_{ph0}$, $\Gamma$, 
    $R_{ph0}$ and $\Delta t$ are plotted in the top-left, top-right, bottom-left and bottom-right
    panels respectively, and color-mapped on the right axis of each panel. 
	The black-dot-dashed line corresponds to $\mathcal{W}$ of the black body spectrum.
	In the first panel, the red-dashed line corresponds to $\mathcal{W}$ of the instantaneous ($\nu\,F_\nu$) spectrum.}}
    \label{fig_width}
\end{figure}

The $\mathcal {W}$ of the order of unity, supported by the observations \citep{axelsson_width}, can be 
obtained only for the case when $\alpha \sim 0.65$ (which is close to 2/3) and shown as yellow triangle in the
{top-left panel} of Figure \ref{fig_width}.
Interestingly, besides providing an explanation for the observed spectral 
width of GRBs, these results also suggest that most of the radiation from the GRB is released during 
the matter-dominated phase of the expanding fireball.  
\section{DISCUSSION AND SUMMARY}\label{sec_results}
The prominence of thermal emission in the GRB prompt phase has been well-studied by various authors \citep{ryde04,ryde2005,peer_08,lundman2013,peer_2011} and 
often supported by observations such as GRB 100724B: \citep{guiriec2013}, GRB 101219B:\citep{ghirlanda_2013}, GRB 081221: \citep{hou2018}, GRB 220426A: \citep{wang2022}. Nevertheless, the prompt emission also carries a significant contribution
of non-thermal spectral component \citep{peer090902B,shabnam_110721,bbzhang_2016,wang2023}. The evolution of the fireball during its coasting phase can deviate the 
spectral shape of the photospheric thermal emission from the Planck function. In this work, we show the spectral shape 
and the peak of the emission from a relativistically expanding fireball are primarily governed by four quantities 
namely, $R_{ph0}$, $T_{ph0}$, $\Gamma$ and $\alpha$ (\S\ref{os}). Besides these, the integration time of the dynamically
evolving fireball can extend the low energy end of the modified thermal emission.

The modified thermal emission obtained from an evolving fireball has a strong dependence on the index $\alpha$.
For lower values of $\alpha$ ($\sim$ 0.1), the spectrum peaks at the high energy corresponding to the photospheric 
temperature. On the other hand, for high values of $\alpha$ ($\gtrsim$ 0.4), it will be governed 
by the 
cooled photon temperature encountered during its evolution (Figure \ref{fig_alp}).
This feature leads to a broad range of 
spectral widths starting from $\sim 0.54$ (nearly equal to that of the Planck function) to $\sim 1.15$. Consistently,
the observed $\mathcal {W}$ can be extrapolated to infer the
energetics of the burst ($\eta \ll 1$) \citep{piran_shemi_naray,axelsson_width}. 
The model can also be applied to individual bursts to estimate the contribution of the thermal component and can 
probe the energetics of the burst.

The choice of other parameters $R_{ph0}$, $T_{ph0}$ and $\Gamma$ will affect only in shifting the spectral peak 
but have minimal impact on the spectral width {(Figure \ref{fig_var} and \ref{fig_width})}. For a given $\alpha$ ($\gtrsim 0.2$), the 
$\mathcal {W}$ will increase with $\Delta t$ only till a certain value. Beyond this, 
the evolution will modify the flux below the half maximum and hence $\mathcal {W}$ will remain constant (In {bottom-right panel, figure \ref{fig_width}}).
\cite{axelsson_width} showed that the distribution of $\mathcal {W}$ for a large sample of GRBs, suggests the 
most probable width to be $\sim 1$. Incidentally, this width closely matches the width of the single
particle emissivity of the synchrotron process. Hence, the authors concluded the GRBs with $\mathcal {W}$ > 1 to be dominated
by the synchrotron process; while the ones with $\mathcal {W}$ < 1 should be consistent with the modified thermal process. However,
\cite{burgess_width} showed the synchrotron process itself is capable to explain many GRBs irrespective of their $\mathcal {W}$ 
estimated from the Band function. In our work, we show that the modified thermal spectrum obtained from the 
evolving fireball is capable of explaining the spectral widths over the range $0.54\lesssim \mathcal {W} \lesssim 1.2$.

If the decoupling of photons happens well within the radiation-dominated phase ($\eta>1$), the temperature remains
nearly constant whereas $\Gamma$ increases with the evolving radius of the fireball. This scenario is not
considered in the present work and has the potential to produce a spectrum which differs from the Planck function.
Though the temperature remains constant, the enhancement in $\Gamma$ will shift the emission to high energies 
resulting in a modified thermal spectrum. Interestingly, the evolved photon temperatures measured by the observer 
will be hotter than the photospheric temperature in such a case. A detailed study under this scenario
will be presented elsewhere. In this work, we have considered only the thermal contribution
of the fireball. Besides this, the burst spectra may involve the non-thermal
emission arising from the shock accelerated particle distribution. The later
contribution is evident from the observed polarisation of the GRBs \citep{chattopadhyay21_grb,Kole20polar_catalog}. Nevertheless, the observed peak spectral
width suggests the thermal origin to be the dominant emission process.

{The authors acknowledge the referee for the valuable suggestions. 
The authors also like to thank Dipankar Bhattacharya and A.R. Rao for the insightful
discussions.}


\bibliography{mbb_vf}{}

\begin{thebibliography}{}
\expandafter\ifx\csname natexlab\endcsname\relax\def\natexlab#1{#1}\fi
\providecommand{\url}[1]{\href{#1}{#1}}
\providecommand{\dodoi}[1]{doi:~\href{http://doi.org/#1}{\nolinkurl{#1}}}
\providecommand{\doeprint}[1]{\href{http://ascl.net/#1}{\nolinkurl{http://ascl.net/#1}}}
\providecommand{\doarXiv}[1]{\href{https://arxiv.org/abs/#1}{\nolinkurl{https://arxiv.org/abs/#1}}}

\bibitem[{{Abramowicz} {et~al.}(1991){Abramowicz}, {Novikov}, \& {Paczynski}}]{abramowic1991}
{Abramowicz}, M.~A., {Novikov}, I.~D., \& {Paczynski}, B. 1991, \apj, 369, 175, \dodoi{10.1086/169748}

\bibitem[{{Amati} {et~al.}(2002){Amati}, {Frontera}, {Tavani}, {in't Zand}, {Antonelli}, {Costa}, {Feroci}, {Guidorzi}, {Heise}, {Masetti}, {Montanari}, {Nicastro}, {Palazzi}, {Pian}, {Piro}, \& {Soffitta}}]{amati02}
{Amati}, L., {Frontera}, F., {Tavani}, M., {et~al.} 2002, \aap, 390, 81, \dodoi{10.1051/0004-6361:20020722}

\bibitem[{{Axelsson} \& {Borgonovo}(2015)}]{axelsson_width}
{Axelsson}, M., \& {Borgonovo}, L. 2015, \mnras, 447, 3150, \dodoi{10.1093/mnras/stu2675}

\bibitem[{{Band} {et~al.}(1993){Band}, {Matteson}, {Ford}, {Schaefer}, {Palmer}, {Teegarden}, {Cline}, {Briggs}, {Paciesas}, {Pendleton}, {Fishman}, {Kouveliotou}, {Meegan}, {Wilson}, \& {Lestrade}}]{band93}
{Band}, D., {Matteson}, J., {Ford}, L., {et~al.} 1993, \apj, 413, 281, \dodoi{10.1086/172995}

\bibitem[{{Beloborodov}(2011)}]{beloborodov_2011}
{Beloborodov}, A.~M. 2011, \apj, 737, 68, \dodoi{10.1088/0004-637X/737/2/68}

\bibitem[{Beloborodov(2012)}]{Beloborodov12}
Beloborodov, A.~M. 2012, The Astrophysical Journal, 762, 13, \dodoi{10.1088/0004-637x/762/1/13}

\bibitem[{{Bharali} {et~al.}(2017){Bharali}, {Sahayanathan}, {Misra}, \& {Boruah}}]{sahayanathan}
{Bharali}, P., {Sahayanathan}, S., {Misra}, R., \& {Boruah}, K. 2017, \na, 55, 22, \dodoi{10.1016/j.newast.2017.02.004}

\bibitem[{{Bo{\v{s}}njak} {et~al.}(2009){Bo{\v{s}}njak}, {Daigne}, \& {Dubus}}]{Bonjak2009}
{Bo{\v{s}}njak}, {\v{Z}}., {Daigne}, F., \& {Dubus}, G. 2009, \aap, 498, 677, \dodoi{10.1051/0004-6361/200811375}

\bibitem[{{Burgess}(2019)}]{burgess_width}
{Burgess}, J.~M. 2019, \aap, 629, A69, \dodoi{10.1051/0004-6361/201935140}

\bibitem[{{Cavallo} \& {Rees}(1978)}]{cavallo_78}
{Cavallo}, G., \& {Rees}, M.~J. 1978, \mnras, 183, 359, \dodoi{10.1093/mnras/183.3.359}

\bibitem[{Chattopadhyay {et~al.}(2021)Chattopadhyay, Gupta, Gupta, Sharma, Iyyani, Ratheesh, Mithun, Aarthy, Palit, Kumar, \& et~al.}]{chattopadhyay21_grb}
Chattopadhyay, T., Gupta, S., Gupta, S., {et~al.} 2021, Journal of Astrphysics and Astronomy, 42, 20, \dodoi{10.1007/s12036-021-09718-2}

\bibitem[{{Cohen} \& {Piran}(1997)}]{cohen_1997}
{Cohen}, E., \& {Piran}, T. 1997, \apjl, 488, L7, \dodoi{10.1086/310916}

\bibitem[{{Daigne} {et~al.}(2011){Daigne}, {Bo{\v{s}}njak}, \& {Dubus}}]{daigne_2011}
{Daigne}, F., {Bo{\v{s}}njak}, {\v{Z}}., \& {Dubus}, G. 2011, \aap, 526, A110, \dodoi{10.1051/0004-6361/201015457}

\bibitem[{{Ghirlanda} {et~al.}(2013){Ghirlanda}, {Pescalli}, \& {Ghisellini}}]{ghirlanda_2013}
{Ghirlanda}, G., {Pescalli}, A., \& {Ghisellini}, G. 2013, \mnras, 432, 3237, \dodoi{10.1093/mnras/stt681}

\bibitem[{{Ghisellini} \& {Celotti}(1999)}]{ghisellini99}
{Ghisellini}, G., \& {Celotti}, A. 1999, \apjl, 511, L93, \dodoi{10.1086/311845}

\bibitem[{{Goodman}(1986)}]{Goodman_86}
{Goodman}, J. 1986, \apjl, 308, L47, \dodoi{10.1086/184741}

\bibitem[{{Guiriec} {et~al.}(2013){Guiriec}, {Daigne}, {Hasco{\"e}t}, {Vianello}, {Ryde}, {Mochkovitch}, {Kouveliotou}, {Xiong}, {Bhat}, {Foley}, {Gruber}, {Burgess}, {McGlynn}, {McEnery}, \& {Gehrels}}]{guiriec2013}
{Guiriec}, S., {Daigne}, F., {Hasco{\"e}t}, R., {et~al.} 2013, \apj, 770, 32, \dodoi{10.1088/0004-637X/770/1/32}

\bibitem[{{Hou} {et~al.}(2018){Hou}, {Zhang}, {Meng}, {Wu}, {Liang}, {L{\"u}}, {Liu}, {Liang}, {Lin}, {Lu}, {Huang}, \& {Zhang}}]{hou2018}
{Hou}, S.-J., {Zhang}, B.-B., {Meng}, Y.-Z., {et~al.} 2018, \apj, 866, 13, \dodoi{10.3847/1538-4357/aadc07}

\bibitem[{{Iyyani} {et~al.}(2013){Iyyani}, {Ryde}, {Axelsson}, {Burgess}, {Guiriec}, {Larsson}, {Lundman}, {Moretti}, {McGlynn}, {Nymark}, \& {Rosquist}}]{shabnam_110721}
{Iyyani}, S., {Ryde}, F., {Axelsson}, M., {et~al.} 2013, \mnras, 433, 2739, \dodoi{10.1093/mnras/stt863}

\bibitem[{Kole {et~al.}(2020)Kole, De~Angelis, Berlato, Burgess, Gauvin, Greiner, Hajdas, Li, Li, Pollo, \& et~al.}]{Kole20polar_catalog}
Kole, M., De~Angelis, N., Berlato, F., {et~al.} 2020, Astronomy and Astrophysics, 644, A124, \dodoi{10.1051/0004-6361/202037915}

\bibitem[{{Lazzati} \& {Begelman}(2010)}]{lazzati_2010}
{Lazzati}, D., \& {Begelman}, M.~C. 2010, \apj, 725, 1137, \dodoi{10.1088/0004-637X/725/1/1137}

\bibitem[{{Lazzati} {et~al.}(2011){Lazzati}, {Morsony}, \& {Begelman}}]{lazzati_2011}
{Lazzati}, D., {Morsony}, B.~J., \& {Begelman}, M.~C. 2011, \apj, 732, 34, \dodoi{10.1088/0004-637X/732/1/34}

\bibitem[{{Liang} {et~al.}(2010){Liang}, {Yi}, {Zhang}, {L{\"u}}, {Zhang}, \& {Zhang}}]{liang10}
{Liang}, E.-W., {Yi}, S.-X., {Zhang}, J., {et~al.} 2010, \apj, 725, 2209, \dodoi{10.1088/0004-637X/725/2/2209}

\bibitem[{{Lundman} {et~al.}(2013){Lundman}, {Pe'er}, \& {Ryde}}]{lundman2013}
{Lundman}, C., {Pe'er}, A., \& {Ryde}, F. 2013, \mnras, 428, 2430, \dodoi{10.1093/mnras/sts219}

\bibitem[{{Meng} {et~al.}(2018){Meng}, {Geng}, {Zhang}, {Wei}, {Xiao}, {Liu}, {Gao}, {Wu}, {Liang}, {Huang}, {Dai}, \& {Zhang}}]{meng2018}
{Meng}, Y.-Z., {Geng}, J.-J., {Zhang}, B.-B., {et~al.} 2018, \apj, 860, 72, \dodoi{10.3847/1538-4357/aac2d9}

\bibitem[{{M{\'e}sz{\'a}ros} \& {Rees}(2000)}]{meszaros_2000}
{M{\'e}sz{\'a}ros}, P., \& {Rees}, M.~J. 2000, \apj, 530, 292, \dodoi{10.1086/308371}

\bibitem[{{Paczynski}(1986)}]{paczy_1986}
{Paczynski}, B. 1986, \apjl, 308, L43, \dodoi{10.1086/184740}

\bibitem[{{Pe'er}(2008)}]{peer_08}
{Pe'er}, A. 2008, \apj, 682, 463, \dodoi{10.1086/588136}

\bibitem[{{Pe'er} \& {Ryde}(2011)}]{peer_2011}
{Pe'er}, A., \& {Ryde}, F. 2011, \apj, 732, 49, \dodoi{10.1088/0004-637X/732/1/49}

\bibitem[{{Pe'er} \& {Zhang}(2006)}]{peer_zhang_2006}
{Pe'er}, A., \& {Zhang}, B. 2006, \apj, 653, 454, \dodoi{10.1086/508681}

\bibitem[{{Pe'Er} {et~al.}(2012){Pe'Er}, {Zhang}, {Ryde}, {McGlynn}, {Zhang}, {Preece}, \& {Kouveliotou}}]{peer090902B}
{Pe'Er}, A., {Zhang}, B.-B., {Ryde}, F., {et~al.} 2012, \mnras, 420, 468, \dodoi{10.1111/j.1365-2966.2011.20052.x}

\bibitem[{{Piran} {et~al.}(1993){Piran}, {Shemi}, \& {Narayan}}]{piran_shemi_naray}
{Piran}, T., {Shemi}, A., \& {Narayan}, R. 1993, \mnras, 263, 861, \dodoi{10.1093/mnras/263.4.861}

\bibitem[{{Preece} {et~al.}(2002){Preece}, {Briggs}, {Giblin}, {Mallozzi}, {Pendleton}, {Paciesas}, \& {Band}}]{preece_2002}
{Preece}, R.~D., {Briggs}, M.~S., {Giblin}, T.~W., {et~al.} 2002, \apj, 581, 1248, \dodoi{10.1086/344252}

\bibitem[{{Rees} \& {Meszaros}(1994)}]{rees94}
{Rees}, M.~J., \& {Meszaros}, P. 1994, \apjl, 430, L93, \dodoi{10.1086/187446}

\bibitem[{{Rees} \& {M{\'e}sz{\'a}ros}(2005)}]{rees_2005}
{Rees}, M.~J., \& {M{\'e}sz{\'a}ros}, P. 2005, \apj, 628, 847, \dodoi{10.1086/430818}

\bibitem[{{Ryde}(2004)}]{ryde04}
{Ryde}, F. 2004, \apj, 614, 827, \dodoi{10.1086/423782}

\bibitem[{{Ryde}(2005)}]{ryde2005}
---. 2005, \apjl, 625, L95, \dodoi{10.1086/431239}

\bibitem[{{Sari} \& {Piran}(1997)}]{sari_1997}
{Sari}, R., \& {Piran}, T. 1997, \mnras, 287, 110, \dodoi{10.1093/mnras/287.1.110}

\bibitem[{{Shemi} \& {Piran}(1990)}]{shemi_90}
{Shemi}, A., \& {Piran}, T. 1990, \apjl, 365, L55, \dodoi{10.1086/185887}

\bibitem[{{Tavani}(1996)}]{Tavani1996}
{Tavani}, M. 1996, \apj, 466, 768, \dodoi{10.1086/177551}

\bibitem[{{Thompson}(1994)}]{Thompson_1994}
{Thompson}, C. 1994, \mnras, 270, 480, \dodoi{10.1093/mnras/270.3.480}

\bibitem[{{Uhm} \& {Zhang}(2014)}]{uhm_14}
{Uhm}, Z.~L., \& {Zhang}, B. 2014, Nature Physics, 10, 351, \dodoi{10.1038/nphys2932}

\bibitem[{{Wang} {et~al.}(2023){Wang}, {Xia}, {Zheng}, {Ren}, \& {Fan}}]{wang2023}
{Wang}, Y., {Xia}, Z.-Q., {Zheng}, T.-C., {Ren}, J., \& {Fan}, Y.-Z. 2023, arXiv e-prints, arXiv:2303.11083, \dodoi{10.48550/arXiv.2303.11083}

\bibitem[{{Wang} {et~al.}(2022){Wang}, {Zheng}, \& {Jin}}]{wang2022}
{Wang}, Y., {Zheng}, T.-C., \& {Jin}, Z.-P. 2022, \apj, 940, 142, \dodoi{10.3847/1538-4357/aca017}

\bibitem[{{Yonetoku} {et~al.}(2004){Yonetoku}, {Murakami}, {Nakamura}, {Yamazaki}, {Inoue}, \& {Ioka}}]{yonetoku04}
{Yonetoku}, D., {Murakami}, T., {Nakamura}, T., {et~al.} 2004, \apj, 609, 935, \dodoi{10.1086/421285}

\bibitem[{{Zhang} \& {Yan}(2011)}]{zhang11}
{Zhang}, B., \& {Yan}, H. 2011, \apj, 726, 90, \dodoi{10.1088/0004-637X/726/2/90}

\bibitem[{{Zhang} {et~al.}(2016){Zhang}, {Uhm}, {Connaughton}, {Briggs}, \& {Zhang}}]{bbzhang_2016}
{Zhang}, B.-B., {Uhm}, Z.~L., {Connaughton}, V., {Briggs}, M.~S., \& {Zhang}, B. 2016, \apj, 816, 72, \dodoi{10.3847/0004-637X/816/2/72}

\bibitem[{{Zhang} {et~al.}(2014){Zhang}, {Chen}, \& {B{\"o}ttcher}}]{zhang14}
{Zhang}, H., {Chen}, X., \& {B{\"o}ttcher}, M. 2014, \apj, 789, 66, \dodoi{10.1088/0004-637X/789/1/66}

\end{thebibliography}
\bibliographystyle{aasjournal}

\end{document}